\documentclass[aps,preprint,showpacs,preprintnumbers,amsmath,amssymb,prd]{revtex4}
\pdfoutput=1
\usepackage{amssymb}
\usepackage{graphicx}

\usepackage{graphics,psfrag}
\usepackage{graphicx,psfrag}
\usepackage{epic,eepic}
\usepackage{color}
\usepackage{epsfig}
\usepackage{latexsym}
\usepackage{pstricks}

\def\be{\begin{equation}}
\def\ee{\end{equation}}
\def\ba{\begin{eqnarray}}
\def\ea{\end{eqnarray}}

\begin{document}

\title{Light bending by nonlinear
electrodynamics under strong electric and magnetic field}

\author{Jin Young Kim }
\email{jykim@kunsan.ac.kr}

\author{Taekoon Lee}
\email{tlee@kunsan.ac.kr}

\affiliation{Department of Physics, Kunsan National University,
Kunsan 573-701, Korea}

%\date{\today}

\begin{abstract}

We calculate the bending angles of light under the strong electric
and magnetic fields by a charged black hole and a magnetized neutron
star according to the nonlinear electrodynamics of Euler-Heisenberg
interaction. The bending angle of light by the electric field of
charged black hole is computed from geometric optics and a general
formula is derived for light bending valid for any orientation of
the magnetic dipole. The astronomical significance of the light
bending by magnetic field of a neutron star is discussed.

\end{abstract}


\maketitle

\section{Introduction}

A nonuniform electric or magnetic background field can induce a
continually varying index of refraction. At classical level the
linearity of the electrodynamics precludes the bending of light by
electric or magnetic field. Therefore any bending must involve a
nonlinear interaction from quantum correction. Because of the
nonlinear optical properties by Euler-Heisenberg interaction
\cite{eulhei,schwinger} arising from the box diagram of quantum
electrodynamics, the light can bend when it passes the neighborhood
of an electrically or magnetically charged object.

Since the maximum available field in a laboratory is of the order
$B,~E/c \sim 10^2 \rm T$, it is difficult to observe the bending
directly in a laboratory
\cite{iacopini,bakalov,boerholten,Piazza06,kpk}. An alternative way
one can think of is observing the effect in the astronomical scale.
There are objects, at least considered theoretically,
 with electric or magnetic field strong enough to make
such bending relevant. One example of light bending by an electrically
charged object is when a high energy photon passes around a charged
 black hole with impact parameter greater than the
Schwarzschild radius \cite{lorenci01}. Another example of light
bending by a magnetic field is when a photon passes the
magnetosphere of a magnetized neutron star with extremely strong
magnetic field of the order $10^{8} - 10^{11} \rm T$. Since neutron
stars have a very dense magnetosphere, only high energy photons may
be observable. Recently the vacuum effect of non-linear
electrodynamics under strong magnetic field by magnetized stars has
been studied widely
\cite{shaviv,heyl99,denisov,denisov01,heylshav02,denisov03,heylshav03,
denisov04,denisov05,heylshav05,dupays,denisov07,heyl10}.

We will calculate the bending angle of the incident photon under the
electric and magnetic field. In the previous work of the authors, we
set up a simple formalism to find the bending angle and the
trajectory of light in a Coulombic field at atomic scale \cite{lbc}.
In this paper,
 we consider the bending of a high energy photon when it passes near
an astronomical object with strong electric or magnetic field. The
organization of the paper is as follow. In Sec. II, we set up the
trajectory equation to calculate the bending angle based on
Euler-Heisenberg interaction. In Sec. III, we consider the bending
by the electric field of a spherically symmetric charged object. In
Sec. IV, we consider the bending by the magnetic field of a magnetic
dipole. We derive a general formula valid for any orientation of the
dipole axis and discuss some special cases where the dipole axis and
the beam  direction are aligned in a particular way. Finally, in
Sec. V, we conclude and discuss the possibility of observation.

\section{The trajectory equation}
The nonlinear interaction of photons is described by the
Euler-Heisenberg Lagrangian \cite{eulhei,schwinger}
 \ba
 {\cal L} &=& -\frac{c^2 \epsilon_0}{4} F_{\mu\nu}F^{\mu\nu} +
 \frac{\alpha^2\hbar^3 \epsilon_0^2}{90m^4c}\left[(
 F_{\mu\nu}F^{\mu\nu})^2+\frac{7}{4} ( F_{\mu\nu}\tilde
 F^{\mu\nu})^2\right]   \nonumber   \\
  &=& \frac{\epsilon_0}{2} ({\bf E}^2 - c^2 {\bf B}^2)  +
 \frac{2\alpha^2\hbar^3 \epsilon_0^2}{45m^4c^5}\left[
 ({\bf E}^2 - c^2 {\bf B}^2)^2 + 7 c^2 ({\bf E} \cdot {\bf B} )^2
 \right].   \label{lagrangian}
 \ea
In the presence of the electric field, the correction to the speed
of light due to the nonlinear interaction is
\cite{Bialynicka,adler,heyher,giesdit,ditgies,heyher98,
lorenci00,lorenci,gies,rikken,lorenci08}

 \be
 \frac{v}{c} = 1-\frac{a\alpha^2\hbar^3 \epsilon_0}{45m^4c^5}
 ({\bf u\times E})^2  ,  \label{velectric}
 \ee
where $\bf u$ denotes the unit vector in the direction of photon
propagation, $a=14$ for the perpendicular mode in which the photon
polarization is perpendicular to the plane spanned by ${\bf u}$ and
${\bf E}$, and $a=8$ for the parallel mode where
 the polarization is parallel to the plane.
Throughout the paper all units are in MKS. For magnetic case, ${\bf
E}$ should be replaced by $c{\bf B}$ in the above equation. Because
the speed of light depends on the electric field the light ray bends
in the presence of a nonuniform field. The bending can be calculated
by geometric optics. The index of refraction due to the background
field is given, in the leading order, by
 \be
 n = \frac{c}{v} = 1 + \frac{a\alpha^2\hbar^3 \epsilon_0}{45m^4c^5}
 ({\bf u\times E})^2 .  \label{nelectric}
 \ee

\begin{figure}
\includegraphics[angle=0, width=6cm ]{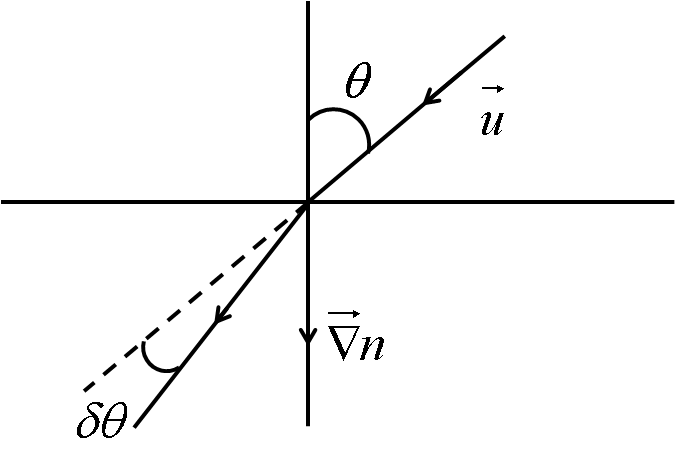} \caption{
Schematic of bending due to nonuniform refractive index. }
\label{fig1}
\end{figure}

The infinitesimal  bending of the photon trajectory
over $\delta\vec{\bf r}$
can be obtained from the Snell's law as
 \be
 \delta \theta =\tan\theta\frac{\delta n}{n}
              =\frac{1}{n}|  \nabla n \times \delta \vec{\bf r}|
 \label{deltheta}
 \ee
where $\delta n={\nabla n}\cdot\delta \vec{\bf r}$ and $\theta$
denotes the angle between ${\bf u}$ and ${\nabla}n$
(see Fig. \ref{fig1}). Since light bends in general toward the
direction of greater index of refraction we can write the bending in
a vector form
 \ba
 \delta {\bf u}=\frac{1}{n}(\delta \vec{\bf r} \times {\nabla}n)\times {\bf u}\,,
 \ea
 which leads to the trajectory equation
 \ba
 \frac{d{\bf u}}{ds}=\frac{1}{n}({\bf u}\times {\nabla}n)\times {\bf u}\,,
 \ea
where $s$ denotes the
distance parameter
 of the light trajectory with
$ds=|d\vec{\bf r}|$ and
 \be
 {\bf u}=\frac{d\vec{\bf r}}{ds}\,.
 \ee
 Since the correction to the index of refraction is generally small in our
consideration the trajectory equation can be approximated to the
leading order as
 \ba
 \frac{d{\bf u}}{ds}=({\bf u}_0\times {\nabla}n)\times {\bf u}_0\,,
 \ea
 where ${\bf u}_0$ denotes the initial direction of the incoming photon.
Throughout the paper we shall assume the photon comes in from
$x=-\infty$ and moves to $+x$ direction, hence
 \be
 {\bf u}_0=(1,0,0)\,,
 \ee
 and putting ${\nabla}n=(\eta_1,\eta_2,\eta_3)$ the trajectory equation becomes
 \ba
 \frac{d^2x}{ds^2}=0\,,\quad \frac{d^2y}{ds^2}=\eta_2\,, \quad
 \frac{d^2z}{ds^2}=\eta_3 \,.
 \ea
 The first equation shows that
$ds=dx$ at leading order, which is obviously true,
 and therefore
the trajectory equations for $y(x)$ and $z(x)$ in the perpendicular
 directions to the incoming photon are
given by
 \ba
 \frac{d^2y}{dx^2}=\eta_2\,,\quad
 \frac{d^2z}{dx^2}=\eta_3\,, \label{trajectory}
 \ea
 which will be used in the following analysis.

\section{Bending by spherically symmetric charged object}
We now consider the bending of photon trajectory by a spherically
symmetric charged object of total charge $Q$. For this case the
bending angle can be calculated in the same way as in Coulombic
case with the electric
field
 \be
 {\bf E}=\frac{Q}{4\pi \epsilon_0 r^2} \hat{r}.
 \ee
For a photon trajectory $y(x)$ moving to $+x$ axis in the $xy$
plane, choose the unit vector in the direction of photon propagation
as
 \be
 {\bf u} = \frac{1}{\sqrt{1 + y'^2 }} (1, y', 0),
 \ee
where prime is the derivative with respect to $x$. The index of
refraction can be written as
 \be
 n = 1 + \frac{a\alpha^2\hbar^3 Q^2}{720 \pi^2 \epsilon_0 m^4c^5}
 \frac{(y-xy')^2}{r^6(1+y'^2)}.
 \ee

\begin{figure}
\includegraphics[angle=0, width=10cm ]{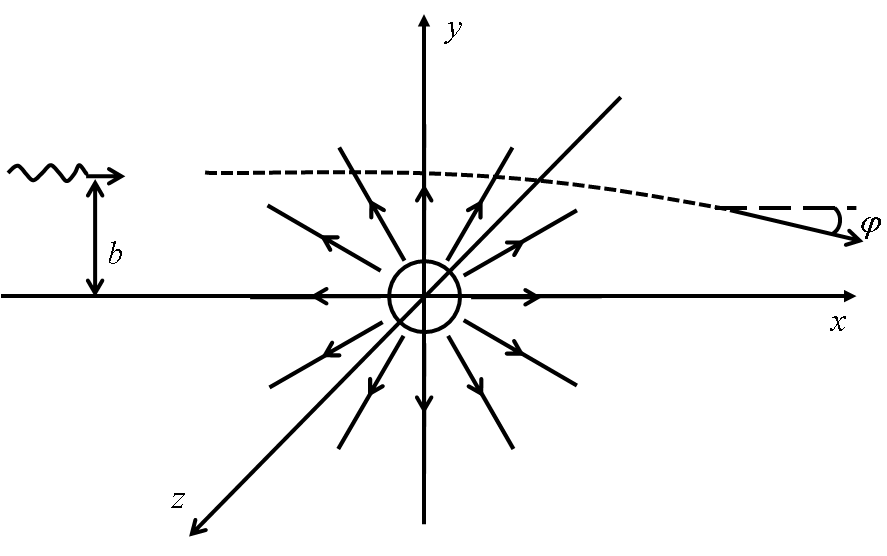} \caption{
Schematic of light bending in a Coulombic field. } \label{fig2}
\end{figure}

For a photon incoming from $x=-\infty$ with impact parameter $b$
(see Fig. \ref{fig2}), the initial condition reads
 \be
 y(-\infty)=b, ~~~~ y'(-\infty)=0 , \label{inicond}
 \ee
and from the first of the trajectory equation (\ref{trajectory})
 \be
 y''=\eta_2 = \frac{a\alpha^2 Q^2 {\lambda}_e^4}
 {360\pi^2 \epsilon_0 \hbar c}(\frac{y}{r^6}-\frac{3y^3}{r^8}),
 \label{trajectoryeltric}
 \ee
 where $\lambda_e = \hbar/mc$ is the Compton length of the electron.
The total bending angle $\varphi_e$ can be obtained by integration
 \be
  y'(\infty) = \int_{-\infty}^{\infty} \eta_2 dx = \tan \varphi_e  \simeq
 \varphi_e
 \ee
  By putting $y=b$ in $\eta_2$, for the leading order solution, we obtain
 \be
 \varphi_{\rm e}
  = - \frac{a\alpha^2 Q^2}{640\pi \epsilon_0 \hbar c}
 \left( \frac{{\lambda}_e}{b} \right)^4 .  \label{bendingelectric}
 \ee
 The bending always occurs toward the center of the charged object
as in the bending by gravitational field.

To compare the bending by electric field with the bending by
gravitation, let us consider a charged non-rotating black hole with
mass ${\cal M}$ and charge $Q$. The bending by gravitational field
is well-known
 \be
 \varphi_{\rm g} = \frac{4G {\cal M}}{bc^2}.
 \ee
 Note that, from $\varphi_{\rm g} \propto 1/b$ and $\varphi_{\rm e} \propto
 1/b^4$, the bending by electric field can be important at short
 distance. The charge and angular momentum per unit mass($J/{\cal M}$)
 is constrained by the mass of the black hole, in Planck units, as \cite{mtw}
 \be
 Q^2 + (J/{\cal M} )^2 \le {\cal M}^2.
 \ee
For non-rotating ($J=0$) charged black
hole, restoring the physical constants, the total electric charge is
constrained by the condition
 \be
 \frac{Q^2}{4 \pi \epsilon_0} \le G {\cal M}^2.
 \ee
We can parameterize the charge as
 \be
 Q =\sqrt{4 \pi \epsilon_0 G} {\cal M} \xi,
\label{charge}
 \ee
 with $0 \le \xi \le 1$. Then the magnitude of the bending
 angle by electric field can be written as
 \be
 \varphi_{\rm e}
  = \frac{a\alpha^2 \xi^2 }{160 }
  \frac{ G{\cal M}^2 }{ \hbar c} \left( \frac{{\lambda}_e}{b} \right)^4
  = \frac{a\alpha^2 \xi^2 }{640 }
  \frac{ bc {\cal M} }{ \hbar } \left( \frac{{\lambda}_e}{b} \right)^4
  \varphi_{\rm g}.
\label{b_eg}
 \ee

To estimate the size of the bending, let us compare the two bending angles for the
maximally charged (extremal, $\xi = 1$) stellar black hole of ten
solar mass ${\cal M}= 10 {\cal M}_{\rm sun}= 2 \times 10^{31} \,{\rm
kg}$. Since our formalism is based on flat space time, not on
general relativity, the impact parameter should be large enough. We
consider the case when the impact parameter is ten times the
Schwarzschild radius, $b= 10 r_{\rm sh} \sim 300 \,{\rm km}$, at
which
 \ba
 \varphi_{\rm g} &=& 1.98 \times 10^{-1} {\rm rad} ;  \nonumber \\
 \varphi_{\rm e} &=& 5.47 \times 10^{1}  \varphi_{\rm g}
 = 1.08 \times 10^{1} {\rm rad} ~({\rm for ~the ~perpendicular~ mode} ~a=14).
 \ea
The bending by electric charge dominates the gravitational bending.
Even for non-extremal charged black hole with $\xi = 0.1$, the
electrical bending, $\varphi_{\rm e} = 1.08 \times 10^{-1} {\rm
rad}$, is comparable to the gravitational bending.

%Because the Euler-Heisenberg Lagrangian is a low-energy effective
%action of QED presented by asymptotic series, the application is
%limited to a weak field approximation. Thus, our formalism can be
%applied only to the region where the field strength is not as strong
This example shows that the bending by electric field can be large.
However, it may not survive the screening by electron-positron
pair creations of
the strong electric field.
As well known, as the electric field strength approaches
the QED critical field
\be
E_{\rm c} = m^2c^3/e \hbar = 1.32 \times
10^{18} {\rm V/m}
\ee
 the electric field is susceptible
to electron-positron
 pair creation.
 In the region where the electric field is of
the order or higher than $E_{\rm c}$, the vacuum is
unstable and the electric field is highly screened by
the pair creation \cite{schwinger,itzykson}. Only
photons entering the region with electric field below $E_{\rm c}$
can have a chance to be observed.

%Let us estimate the field strength for the case $\xi =0.1$ and $b =
%10 r_{\rm sh}$ where the electrical bending has the same order as
%the gravitational bending. Since the electric field at a distance
%$b$ is given by
% \be
% E = \sqrt{\frac{G}{ 4 \pi \epsilon_0}} \frac {{\cal M} \xi} {b^2},
%  \label{electricfield}
% \ee
%the field strength at $ b = 10 r_{\rm sh}$ and $\xi = 0.1$ is
%estimated as $1.7 \times 10^{19} {\rm V/m}$. This number is about
%two orders of magnitude above $E_{\rm c}$. Even for $\xi = 0.01$,
%where $\varphi_{\rm e} \lesssim 10^{-2}  \varphi_{\rm g}$, the field
%strength is about order of ten above the critical field. If we
%consider the region where the electric field is below one percent of
%$E_{\rm c}$ to apply our calculation securely, the impact parameter
%should be larger than $10^3 r_{\rm sh}$. At $b = 10^3 r_{\rm sh}$,
%the ratio of two bending angles is estimated as
% $ \varphi_{\rm e} / \varphi_{\rm g} \lesssim 10^{-4}$.
%So we conclude that the bending of light by the electric field of a
%charged black hole is negligibly small compared with the
%gravitational bending in the observable region.
To simplify the discussion we may write (\ref{b_eg}) as
 \be
\frac{\varphi_{\rm e}}{\varphi_{\rm g}} \approx 0.083
\left(\frac{a}{10}\right)\sqrt{\frac{\xi}{\chi}} \left(\frac{E_{\rm
b}}{E_{\rm c}}\right) \left(\frac{b_{\rm c}}{b}\right) \label{ratio}
,
 \ee
where  $\chi$ is the ratio of the black hole mass to the solar mass,
 \be
 \chi=\frac{\cal M}{{\cal M}_{\rm sun}}\,,
 \ee
 $E_{\rm b}$ is
the Coulomb field at radius $b$: \be E_{\rm b}=\frac{Q}{4\pi
\epsilon_0 b^2} \ee with $Q$ given by (\ref{charge}), and $b_{\rm
c}$ denotes the radius at which $E_{\rm b}=E_c$ given by
 \be
 b_{\rm c}\approx 365\sqrt{\frac{\xi}{\chi}}R_{0} ,
 \ee
 where $R_{0}$ is the
Schwarzschild radius of the black hole $R_0=2G{\cal M}/c^2$.
Eq.~(\ref{ratio}) shows that at $b$, for which the electric field
strength is below $E_c$, the ratio is bounded by \be
\frac{\varphi_{\rm e}}{\varphi_{\rm g}} <
 \frac{0.083}{\sqrt{\chi}} \left(\frac{a}{10}\right)\,,
\ee
which shows that the screening by pair creation severely bounds the
valid range of the bending angle by a black hole with mass larger than
the solar mass. The light bending
can be comparable to the gravitational bending only for black holes
with mass substantially smaller than the solar mass.

\begin{figure}
\includegraphics[angle=0, width=10cm ]{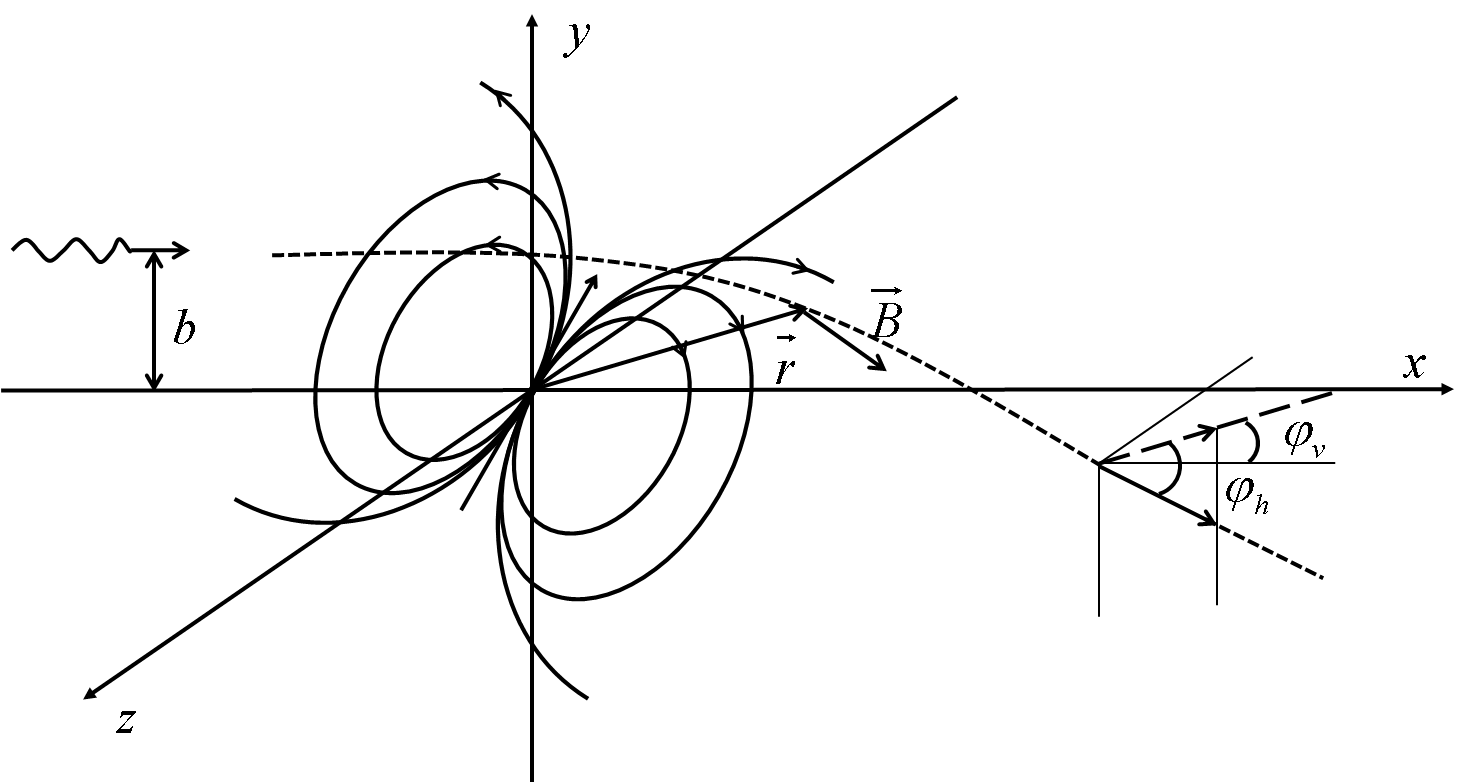} \caption{
Schematic of light bending by magnetic dipole. The magnetic dipole
is located at the origin, the dotted line is the photon trajectory,
the dashed line is the projection of outgoing photon path on the
$xz$-plane, and $\phi_h$ ($\phi_v$) is the bending angle of
horizontal (vertical) direction. } \label{fig3}
\end{figure}

\section{Bending by magnetic dipole}
We consider the bending of photon trajectory by a magnetic dipole.
Obviously, the bending by magnetic dipole
should  depend on the orientation of dipole relative to the direction of
the incoming photon. We consider the bending by a magnetic dipole
located at the origin with arbitrary orientation (see Fig.
\ref{fig3}). We take the direction of the incoming photon as $x$
axis, horizontal direction as $y$ axis, and vertical direction as
$z$ axis. For the magnetic dipole $\bf M$ located at origin, we
define the directional cosines $\alpha = \hat M \cdot \hat {x}$,
$\beta = \hat M \cdot \hat {y}$, $\gamma = \hat M \cdot \hat { z}$
such that ${\bf M} = M {\hat  M }$. The magnetic field by the dipole
is given by
 \be
 {\bf B} =(B_x , B_y , B_z ) = \frac{\mu_0 M}{4 \pi}
 \left( \frac {3( {\hat  M} \cdot {\vec r} ) {\vec r}}{r^5}
 -\frac {\hat M}{r^3} \right),  \label{bfield}
 \ee
and the index of refraction due to this background magnetic field is
 \be
 n = \frac{c}{v} = 1 + \frac{a\alpha^2\hbar^3 \epsilon_0}{45m^4c^3}
 ({\bf u\times B})^2 .  \label{nmagnetic}
 \ee
Taking the unit vector in the direction of photon propagation as
 \be
 {\bf u} = \frac{1}{\sqrt{1 + y'^2 + z'^2}} (1, y' , z'),
 \ee
the index of refraction can be written as
 \be
 n = 1 + \frac{a\alpha^2\hbar^3 \epsilon_0}{45m^4c^3}
 \left( \frac{\mu_0 M}{4 \pi} \right)^2
 \frac{1}{1 + y'^2 + z'^2}
 \left[ (B_z y' - B_y z')^2 + (B_z - B_x z' )^2 + (B_y - B_x y' )^2
 \right]. \label{nmaggeneral}
 \ee

To the leading order, the index of
 refraction is written explicitly
as
 \ba
 n &=& 1 + \frac{a\alpha^2\hbar^3 \epsilon_0}{45m^4c^3}
 \left( \frac{\mu_0 M}{4 \pi} \right)^2
 \frac{1}{r^{10} }
 \Bigg[  \Big\{ \beta(2y^2 - x^2 -z^2) + 3 \alpha xy +
 3 \gamma yz \Big\}^2    \nonumber \\
 &&~~~~~~~~~~~~~~~~~~~~~~~~~~~~~~~~~
 +\Big\{ \gamma (2z^2 - x^2 -y^2) + 3 \alpha xz + 3 \beta yz  \Big\}^2
 \Bigg]\,,
 \label{nmagleading}
 \ea
and from the photon trajectory
equation (\ref{trajectory}) we have
 \ba
 y''=&&
 \frac{2 a\alpha^2\hbar^3 \epsilon_0}{45m^4c^3}
 \left( \frac{\mu_0 M}{4 \pi} \right)^2
 \Bigg[ ~~~\frac{2}{ r^{10} }  \Big\{ \left( \beta(2y^2 - x^2 -z^2) + 3 \alpha
 xy + 3 \gamma yz \right)(4 \beta y +3 \alpha x+3 \gamma z )  \nonumber \\
  &&~~~~~~~~~~~~~~~~~~~~~~~~~~~~~~~~~~+
  \left(\gamma (2z^2 - x^2 -y^2) + 3 \alpha
 xz + 3 \beta yz\right)(-2 \gamma y +3 \beta z ) \Big\} \nonumber \\
  &&~~~~~~~~~~~~~~~~~~~~~~~~~~ -
  \frac{10y}{r^{12} }  \Big\{ \left(\beta(2y^2 - x^2 -z^2)
  + 3 \alpha xy + 3 \gamma yz\right)^2  \nonumber \\
  &&~~~~~~~~~~~~~~~~~~~~~~~~~~~~~~~~~~+
  \left(\gamma (2z^2 - x^2 -y^2) + 3 \alpha xz + 3 \beta yz \right)^2
 \Big \}   \Bigg] ,
 \label{trajectoryhorizontal}
 \ea
 \ba
 z''=&&
 \frac{2 a\alpha^2\hbar^3 \epsilon_0}{45m^4c^3}
 \left( \frac{\mu_0 M}{4 \pi} \right)^2
 \Bigg[ ~~~\frac{2}{ r^{10} }  \Big\{ \left(\beta(2y^2 - x^2 -z^2) + 3 \alpha
 xy + 3 \gamma yz \right)(-2 \beta z +3 \gamma y )  \nonumber \\
  &&~~~~~~~~~~~~~~~~~~~~~~~~~~~~~~~~~~+ \left(\gamma (2z^2 - x^2 -y^2) + 3 \alpha
 xz + 3 \beta yz \right)(4 \gamma z + 3 \alpha x +3 \beta y ) \Big\} \nonumber \\
  &&~~~~~~~~~~~~~~~~~~~~~~~~~~ - \frac{10z}{r^{12} }  \Big\{ \left(\beta(2y^2 - x^2 -z^2)
  + 3 \alpha xy + 3 \gamma yz\right)^2  \nonumber \\
  &&~~~~~~~~~~~~~~~~~~~~~~~~~~~~~~~~~~+
  \left(\gamma (2z^2 - x^2 -y^2) + 3 \alpha xz + 3 \beta yz \right)^2
 \Big \}   \Bigg] .
 \label{trajectoryvertical}
 \ea
The total bending angle can be obtained by integration by putting $y=b$ and $z=0$
 in (\ref{trajectoryhorizontal}) and (\ref{trajectoryvertical}),
which gives for horizontal
($\varphi_h = y'(\infty)$) and vertical ($\varphi_v = z'(\infty)$)
deflections,
 \ba
 \varphi_h
  &=& - \frac{ \pi}{3 \cdot 2^7} \frac{a\alpha^2 \epsilon_0 c}{ \hbar }
 \left( \frac{\mu_0 M}{4 \pi} \right)^2
 \frac{{\lambda}_e^4}{b^6} (15 \alpha^2 +41 \beta^2 +16 \gamma^2),
 \label{totalhorizontalbending} \\
  \varphi_v
  &=&  \frac{5 \pi}{3 \cdot 2^6} \frac{a\alpha^2 \epsilon_0 c}{ \hbar }
 \left( \frac{\mu_0 M}{4 \pi} \right)^2
 \frac{{\lambda}_e^4}{b^6} \beta \gamma. \label{totalverticalbending}
 \ea

An important comment is in order. The result is valid as long as the
polarization of the photon remains, throughout the trajectory,
perpendicular or parallel to the plane spanned by ${\bf u} $ and
${\bf B}$. It can be easily checked that this
 condition is met only when the
magnetic dipole axis is either in $\hat{z}$ direction or in the $xy$
plane. When the dipole axis is in none of these directions the
photon polarization remains neither in pure perpendicular nor in
pure parallel mode, even if it started in one of the modes.
  In this case the birefringence effect takes place,
and the perpendicular and the parallel modes of the light ray split.
Since the splitting occurs continually on every branch-out rays,
this cascade of splitting results in the initial single light ray
branching out to a light bundle. The bending angles
(\ref{totalhorizontalbending}) and (\ref{totalverticalbending}),
with $a$ either 14 for the perpendicular mode or 8 for the parallel
mode, then provide an envelope for the maximal and minimal bending
angles of the light bundle in the horizontal and perpendicular
direction, respectively.

% case i) perpendicular and passing equator

\begin{figure}
\includegraphics[angle=0, width=9cm ]{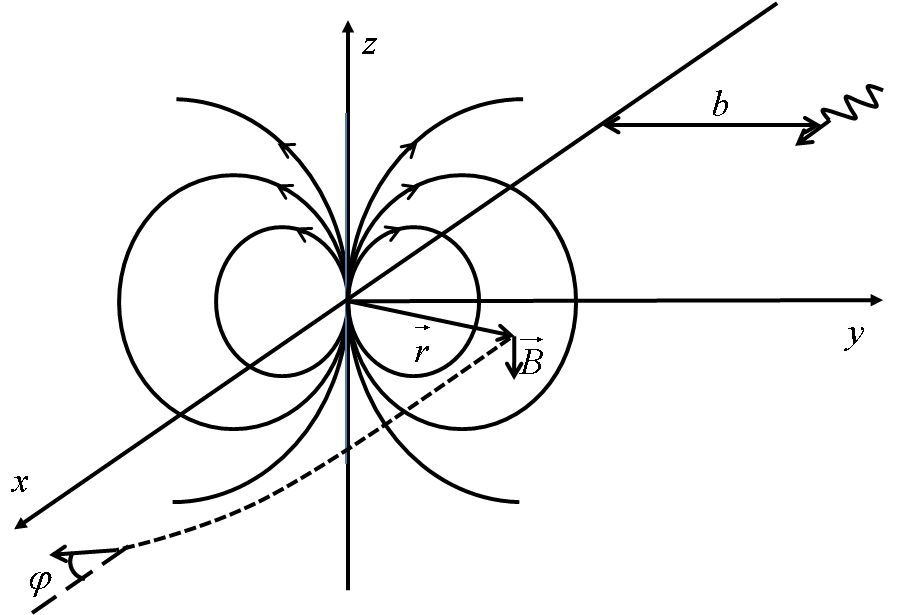} \caption{
Schematic of light bending by magnetic dipole when the photon path
is perpendicular to the dipole moment and traveling on the equator of
the dipole. } \label{fig4}
\end{figure}

Let us now consider some special cases which will allow us to
compare our result with the previous work. First, we consider the
case when the photon path is perpendicular to the  dipole moment and
traveling on the equator of the dipole. Assume that the magnetic
moment directs along $\hat{z}$
 and the incident photon is coming from $x=-\infty$ (see Fig.
\ref{fig4}). This is the specific case considered by Denisov et al.
\cite{denisov}. Taking $\alpha=\beta=0$ and $\gamma=1$, the magnetic
field on the dipole equator ($xy$ plane) is given by
 \be
 {\rm \bf B } = \frac{\mu_0 M}{4 \pi} \frac {1}{r^3} \hat{ z}.
 \ee
By symmetry, there is no vertical ($z$) bending, so $\varphi_v = 0$,
and the horizontal ($y$) bending angle is given by
 \be
 \varphi_h
  = - \frac{ \pi}{24} \frac{a\alpha^2 \epsilon_0 c}{ \hbar }
 \left( \frac{\mu_0 M}{4 \pi} \right)^2
 \frac{{\lambda}_e^4}{b^6}. \label{bendingperp-equator}
 \ee
This result agrees with Denisov et al. (see
the Eqs. (4) and (5) in \cite{denisov}).

% case ii) parallel

\begin{figure}
\includegraphics[angle=0, width=10cm ]{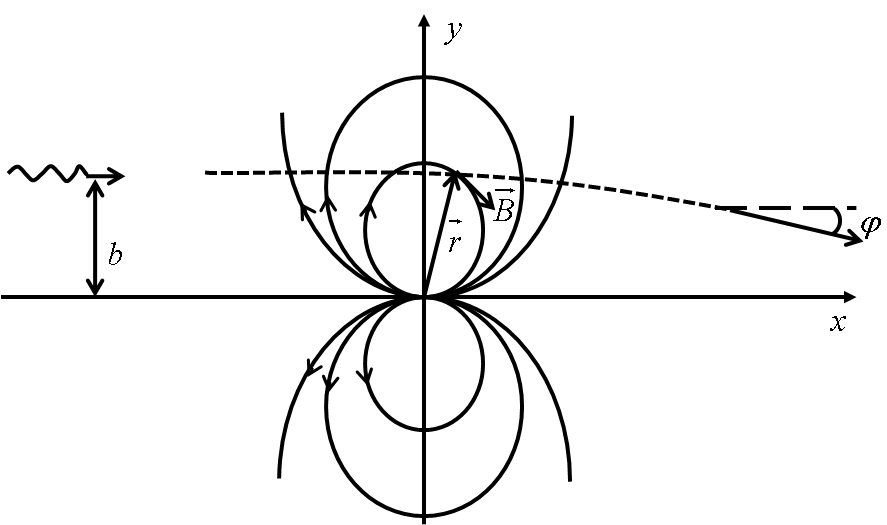} \caption{
Schematic of light bending by a magnetic dipole, located at origin
directing $-x$ direction, when the photon path is parallel to the
dipole } \label{fig5}
\end{figure}

Second, consider the case when the photon path is parallel or
anti-parallel to the dipole axis. Assume that the dipole at the
origin directs along the
 $-x$ axis in the $xy$ plane (see Fig. \ref{fig5}).
 Taking $\alpha=-1$ and $\beta=\gamma=0$, the magnetic field on the
$xy$ plane is given by
 \be
 {\bf B } = \frac{\mu_0 M}{4 \pi} \frac {1}{r^5} \left ( (y^2 -
 2 x^2 ) \hat {x}  -3 xy \hat{y} \right ).
 \ee
Also there is no bending in $z$ direction, $\varphi_v = 0$, and the
bending angle in $y$ direction is given by
 \be
 \varphi_h
  = - \frac{5 \pi}{2^7} \frac{a\alpha^2 \epsilon_0 c}{ \hbar }
 \left( \frac{\mu_0 M}{4 \pi} \right)^2
 \frac{ {\lambda}_e^4}{b^6}. \label{bendingparallel}
 \ee

% case iii) perpendicular and passing north pole

\begin{figure}
\includegraphics[angle=0, width=10cm ]{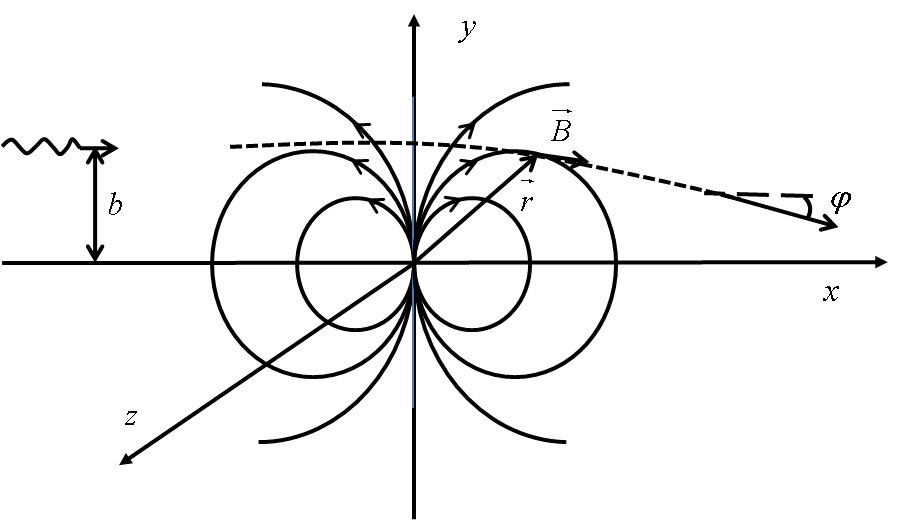} \caption{
Schematic of light bending by magnetic dipole when the photon path
is perpendicular to dipole moment and passing the axis of the
dipole. } \label{fig6}
\end{figure}

Finally, consider the case when the photon path is perpendicular to
dipole moment and passes the north or south pole. Locate the dipole
at the origin directing along the $y$ axis in the $xy$ plane (see
Fig. \ref{fig6}). Taking $\alpha=\gamma=0$ and $\beta=1$, the
magnetic field on the $xy$ plane is given by
 \be
 {\bf B } = \frac{\mu_0 M}{4 \pi} \frac{1}{r^5} \left( 3xy \hat{ x}
  + (2y^2 -x^2) \hat{ y} \right).
 \ee
 There is also no vertical bending, $\varphi_v
= 0$, and the horizontal bending angle is given by
 \be
 \varphi_h
  = - \frac{ 41 \pi}{3 \cdot 2^7} \frac{a\alpha^2 \epsilon_0 c}{ \hbar }
 \left( \frac{\mu_0 M}{4 \pi} \right)^2
 \frac{{\lambda}_e^4}{b^6}. \label{bendingperp-np}
 \ee
 This configuration gives the maximum possible bending;
 The gradient of index of refraction is
 maximal along this direction.

Let us now compare the bending by magnetic
field with the gravitational bending.
 We will consider the possible maximum bending given by Eq.
(\ref{bendingperp-np}) for a strongly magnetized neutron star with
solar mass $ {\cal M}= {\cal M}_{\rm sun} = 2 \times 10^{30}\, {\rm kg}$ and
radius $r_0 = 10 \,{\rm km}$. For extremely strongly magnetized
neutron stars the magnetic field at the surface can be strong as
$B_{\rm s} = 10^8 - 10^{11} {\rm T}$. Parameterizing the impact
parameter in units of the radius $b =\zeta r_0 $ with $\zeta
>1$, the bending by a magnetic field can be written as
 \be
 \varphi_{\rm m}
  %= \frac{ 41 \pi}{3 \cdot 2^7} \frac{a\alpha^2 \epsilon_0 c}{ \hbar }
 %\left( \frac{\mu_0 M}{4 \pi r_0^3} \right)^2
 %\frac{{\lambda}_e^4}{\zeta^6}
  =
 \frac{ 41 \pi}{3 \cdot 2^7} \frac{a\alpha^2 \epsilon_0 c}{ \hbar }
 B_{\rm s}^2 \frac{{\lambda}_e^4}{\zeta^6},
 \label{bendinmagmax}
 \ee
 where we have used $B_{\rm s} = {\mu_0 M}/{4 \pi r_0^3}$,
 the magnetic field strength at
 the neutron star surface.
 For the magnetic field strength of $B_s = 10^9 \rm T$ of a neutron star,
 the maximum value of the bending angle is given by
 \be
 \varphi_{\rm m} = 1.40 \times 10^{-4} ~{\rm rad}\,,
 \ee
 which is for a ray glancing the north or
south pole where the bending is maximal. This is much smaller than the
gravitational bending $\varphi_{\rm g}=0.59 ~{\rm rad}$.

By increasing the surface magnetic field  a larger bending angle can
be obtained. However, as in the bending by charged black hole, the
bending is constrained by the screening of electron-positron pair
creation of strong magnetic field above the critical strength
 \be
B_{\rm c}=\frac{m^2 c^2}{e\hbar}=4.4\times 10^9 ~{\rm T}\,.
 \ee
To see this  we may write  (\ref{bendinmagmax}) as
 \be
 \frac{\varphi_{\rm m}}{\varphi_{\rm g}}\approx 10^{-3}
 \left(\frac{a}{10}\right) \left(\frac{b}{R_0}\right)
 \left(\frac{B_{\rm b}}{B_{\rm c}}\right)^2\,,
 \ee
 where $B_{\rm b}$ is the magnetic field at radius $b$ given
 by
 \be
 B_{\rm b} =B_{\rm s} \left(\frac{r_0}{b}\right)^3\,,
 \ee
 and $R_0$ denotes the Schwarzschild radius
 $R_0=2G{\cal M}/c^2$. To avoid the screening we require
 $B_{\rm b}< B_{\rm c}$, which gives the bound
 \be
 \frac{\varphi_{\rm m}}{\varphi_{\rm g}}< 10^{-3}
 \left(\frac{a}{10}\right) \left(\frac{b_{\rm c}}{R_0}\right)\,,
 \ee
 where $b_{\rm c}$ denotes the radius at which $B_b=B_c$.
 Thus,  a larger bending can be obtained with
 large $b_{\rm c}/R_0$, but it is obviously bounded
 by  the surface
 magnetic field strength through the relation
 \be
 B_{\rm s}=B_{\rm c}\left(\frac{b_{\rm c}}{r_0}\right)^3\,.
 \ee
 For a magnetar with $B_{\rm s}=100 B_{\rm c}$, for example, we get
 $b_{\rm c}=4.6 r_0$, and considering that $r_0/R_0\sim {\cal O}(1)$
 the magnetic bending is expected to be, at most, a few
 percent of the gravitational bending.

\section{Discussion}
We have studied how  photons can be bent when they travel through
the strong electric or magnetic field of compact object like a
charged black hole or a neutron star. We calculated the bending
angles according to the nonlinear electrodynamics of
Euler-Heisenberg interaction. Our calculation shows that the bending
by electrically charged astronomical object can be comparable or
larger than the gravitational bending. However,
 the screening by the electron-positron
pair creation strongly bounds the electric bending so that
the electric bending can be significant compared with the
gravitational bending  only for black holes with mass substantially
smaller than the solar mass. We also found a general formula for
light bending by magnetic field,
valid for any orientation of the magnetic dipole. Our calculation
shows that for a magnetar with surface magnetic field
$B_{\rm s}=10^{11} ~{\rm T}$
the magnetic bending can be, at most, a few percent of the gravitational
bending.

%So far laser interferometers are the popular apparatus to observe
%the nonlinear properties of the QED. For example,
%fluctuation-induced diffractive effects induced by high-intensity
%laser pulses have recently been discussed by King et al. \cite{kpk}.
%They presented a matterless double-slit scenario to observe a
%photon-photon scattering and demonstrate the possibility of both
%controlling light with light and non-locally investigating features
%of the quantum vacuum structure.

Since the magnetic bending is expected to be small compared to
gravitational bending any chance of observation may be realized only
when the experiment detects the small variation over the
gravitational bending.
One way to observe the light
bending by magnetic field may be using the birefringence
\cite{shaviv} that the bending of perpendicular polarization is
1.75(=14/8) times larger than the bending of parallel polarization.
Even in the region where the bending by magnetic field is weak
compared with the gravitational bending, by eliminating the overall
gravitational bending, the polarization dependence of the bending by
magnetic field may be tested if the allowed precision is sufficient
enough.

Another possibility is detecting the time variation in bending.
The magnetic axis  of a magnetar is likely to be different from
 the rotational axis. In this case the magnetic
  bending by a rapidly spinning
 magnetar will add a small wiggling effect over the gravitational
  bending.

\begin{figure}
\includegraphics[angle=0, width=10cm ]{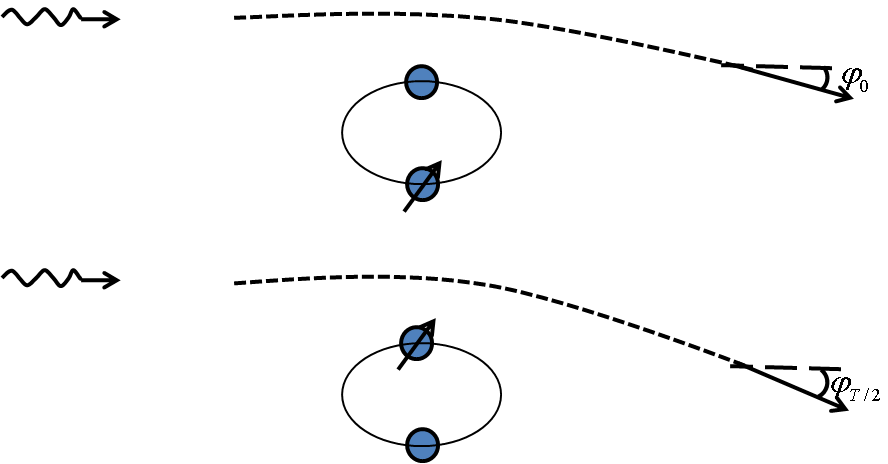} \caption{
Schematic of detecting the light bending by a neutron star and its
binary system of equal mass. The bending angles at $t=0$ and $t =
T/2$ can be different by the contribution of magnetic field of the
neutron star. } \label{fig7}
\end{figure}

The neutron star in a binary system \cite{dupays} may also be used
to detect the magnetic bending. Most of the
neutron stars are isolated stars, less than one hundred are known to
be in binary systems with nondegenerate stars. Assume for simplicity
that the nondegenerate companion star has the same mass as the
neutron star. If we consider only the bending by gravitation, the
bending at $t=0$ and the bending at a half orbital period later
$t=T/2$ will be the same. However, when the bending by magnetic
field is included, the bending angles will be different by the
relative position of the two stars (see Fig. \ref{fig7}).

\acknowledgements{ We would like to thank Y. Yi, M. I. Park, and M.
K. Park for useful discussions. This paper was supported by research
funds of Kunsan National University. }

\end{document}